\documentclass[aps,preprint,preprintnumbers,amsmath,amssymb,floatfix,showpacs]{revtex4}
\usepackage{graphicx}
\usepackage{epsfig}

\usepackage{amsbsy,bm}

\begin{document}

\title{The hybrid surface flux method for the ionization amplitude extraction from the calculated wavefunction}

\author{Vladislav V. Serov and Tatyana A. Sergeeva}

\affiliation{Department of Theoretical Physics, Saratov State University, 83 Astrakhanskaya, Saratov 410012, Russia}

\author{Sergue I. Vinitsky}

\affiliation{Bogoliubov Laboratory of Theoretical Physics, Joint Institute for Nuclear Research, Dubna 141980, Russia}

\date{\today}

\begin{abstract}
A method is proposed for extracting ionization amplitudes from the solution of the time-dependent Schr\"odinger equation (TDSE) describing a system in a time-dependent external field.  The method is a hybrid of two earlier developed methods, the time-dependent surface flux (t-SURFF) method and the method using the propagated wavepacket as the source term in a time-independent driven Schr\"odinger equation with the field-free Hamiltonian. It is demonstrated that the method combines the advantages of the parent ones and allows the extraction of ionization amplitudes by solving the TDSE within a small spatial domain (with the boundary conditions provided by exterior complex scaling) and a time interval, not exceeding the external field pulse duration.
\end{abstract}



\maketitle

\section{Introduction}
Nowadays the attosecond processes in atoms and molecules triggered by the ultra-short very-high-power laser pulses are extensively investigated. Their theoretical description requires the numerical solution of the time-dependent Schr\"odinger equation (TDSE) for a system containing several particles. The key issue of the numerical methods used for this purpose is the following extraction from the calculated wavefunction of the information about the experimentally observed quantities such as the energy and angular distribution of the ejected electrons.

For the aim of the ionization amplitudes obtaining from the wavefunction a number of approaches has been proposed. Among them one should mention those rest upon the evaluation of the projection to the approximate continuum wave function \cite{Telnov2009,Thumm2003}, the space Fourier transform of the wavefunction asymptotic part \cite{Toshima2006}, the temporal Fourier transform (tFT) of the wavefunction \cite{BouckeKull1997,Dondera2010} or the autocorrelation function \cite{Nikolopoulos2007}. 
The latter approach has also the refined version, namely the technique built upon the tFT of the probability amplitude flux through a certain closed surface \cite{Selin1999,Serov2001,Serov2011,Tao2012,Scrinzi2012,OurRev2013} (following \cite{Tao2012}, hereinafter we will refer to this method as the t-SURFF). Finally, McCurdy et al. have suggested an approach rest upon the evaluation of the flux of the probability amplitude for the scattering function derived by the Green's function operator action to the wavefunction after the end of the pulse \cite{McCurdy2007,McCurdyRev2013} (in what follows we will refer to this method as the E-SURFF). The drastic solution to the problem under consideration is the utilizing of the time-dependent scaling approach \cite{Kaschiev2003,Serov2007,Serov2008,Serov2010,Piraux2011}.  

However, all of the above itemized schemes have various drawbacks. For example, the time-dependent scaling approach \cite{Kaschiev2003,Serov2007} faces with difficulties due to the inadequate bound states modelling. Next, the E-SURFF \cite{McCurdy2007,McCurdyRev2013} method implementation requires the enormous space region in order not to allow the ejected electron wave packet to reach its boundaries before the end of the laser pulse action. At the same time, the t-SURFF approach \cite{Selin1999,Serov2001,Serov2011,Tao2012} though allowing to utilize rather compact space grid (under the condition of the employing of any methods for the supposition of the electron wavefunction unphysical reflection from the grid boundary, see below), demands the TDSE solving for a time period large enough to provide the probability to ``flow'' outwards the outer closed surface. 

In the present work we propose a method for the amplitudes extraction combining the advantages of both the E-SURFF \cite{McCurdy2007} and the t-SURFF \cite{Selin1999} approaches. The paper structure is as follows. The Section \ref{Sec:NonphysicalReflection} outlines the commonly used methods for the supposition of the unphysical boundary reflection allowing to obtain the TDSE solution on quite moderate space region, in particular the ECS technique. The Sec. \ref{Sec:AmplExtr} presents the theoretical derivation of the proposed approach of the amplitudes exctraction, designated the t\&E-SURFF, and establishes the relation with the advanced t-SURFFc method suggested by the authors earlier \cite{Serov2011}. The Sec. \ref{Sec:NumericalModel} demonstrates the new approach benefits over both the t-SURFF and the E-SURFF by the example of the benchmark one-dimensional problem. The Sec. \ref{Sec:Concl} contains the concluding remarks, particularly, possible problems in the t-SURFFc and t\&E-SURFF utilizing for the double ionization amplitudes evaluation is discussed. In the Appendix the t-SURFF implementation in terms of the length, velocity and acceleration gauges is briefly defined.

\section{Unphysical reflection and the ways for their suppression}\label{Sec:NonphysicalReflection}

A particle motion in the laser field is described under the dipole approximation by the TDSE having in the length gauge the following form:
\begin{eqnarray}
i\frac{\partial\psi(\mathbf{r},t)}{\partial
t}=\left[\hat{H}_0-e\mathbf{E}(t)\cdot\mathbf{r}\right]\psi(\mathbf{r},t).
\label{TDSE3D}
\end{eqnarray}
Here
\begin{eqnarray}
\hat{H}_0=-\frac{1}{2}\nabla^2 + U_0(\mathbf{r})
\label{H0}
\end{eqnarray}
is an unperturbed system Hamiltonian, $\mathbf{E}(t)$ is the external electric field strength, $e$ is the particle charge.

If the system potential  $U_0(\mathbf{r})$ is short-range, then the Eq. \eqref{TDSE3D} reduces far from the center to the TDSE for a free electron in the variable external uniform field. This TDSE solution is well known to be the Volkov function, having the following form in the length gauge:
\begin{eqnarray}
 \Theta_{\mathbf{k}\,}(\mathbf{r},t)=(2\pi)^{-3/2}
\exp\left\{i\left[
\mathbf{p}(t)\cdot\mathbf{r}-\frac{1}{2}\int_0^t p^2(\tau)d\tau
\right]\right\}, \label{VolkovCoord}
\end{eqnarray}
where $\mathbf{k}$ is the conserved canonical momentum, $\mathbf{p}(t)=\mathbf{k}-\frac{e}{c}\mathbf{A}(t)$ is the kinetic momentum, $\mathbf{A}$ denotes the external field vector potential. For the electric field having the strength $\mathbf{E}(t)$ the latter may be written as $\mathbf{A}(t)=-c\int_0^t\mathbf{E}(\tau)d\tau$.

From the complete Volkov functions set one may obtain the Green's function in an analytic form \cite{Selin1999,Tao2012} and thus have an opportunity to completely determine the wavefunction evolution far from the center. Hence the wavefunction evolution appears to be non-trivial only in the vicinity of the center, so the Eq.\eqref{TDSE3D} solving in the enormous region fully enclosing the ejected electrons wave packet seems to be impracticle. However the simple treating of the smaller region under the setting the time-independent boundary conditions (e.g. Dirichlet or Neumann boundary conditions) appears to be not enough. That is, a boundary acts as a mirror reflecting the electrons back into a region of TDSE solution leading to the fully physically inadequate results. This particularity is conventionally referred to as the unphysical reflection.

The mathematically consistent approach to deal with this problem is the utilizing of the integral boundary conditions  \cite{BouckeKull1997,Selin1999,Selin2000}. However, their technical implementation in the many-dimensional case becomes quite cumbersome. Moreover, these boundary conditions require both the memorizing of the wavefunction meanings on the boundary for all the time values as well as the evaluation of an integral over time at the every time step. As a result, the calculation for a large time period becomes very memory-demanding, while the computation time depends quadratically on the simulated physical time duration. 

Another approaches are based upon the idea to absorb or dampen the ejected electron wave packet in order to get an opportunity to set the Dirichlet boundary conditions at the boundary. Among these techniques one may note the imposing of the auxilary imaginary absorbing potential \cite{Sajeev2009}, as well as the multiplying by a mask function \cite{Kulander1992}. But the most efficient scheme of this kind is the so called ECS \cite{McCurdy1991,McCurdyRev2004}. It consists in the rotation of the integration contour over the radial coordinate $r$ beyond a certain point $r_{CS}$ into the complex plane: 
\begin{eqnarray}
 r\to z=\left\{
 \begin{array}{ll}
 r,& r\in [0,r_{CS}]; \\
 r_{CS}+e^{i\theta_{CS}}(r-r_{CS}),& r>r_{CS},
 \end{array} \right. \label{ECSdef}
\end{eqnarray}
where $\theta_{CS}$ is the angle of the contour rotation. As a result, the functions describing the outgoing waves attain the exponentially damped envelops
\begin{eqnarray*}
|\exp(ip_rz)|=\exp[-p_r(r-r_{CS})\sin\theta_{CS}].
\end{eqnarray*}
Provided that the scaling region $l_{CS}=r_{max}-r_{CS}$ is chosen to be large enough, the envelops on the grid boundary become negligible, so that one becomes permitted to utilize the Dirichlet boundary conditions. However, since the complex scaling poorly dampens the waves corresponding to the electrons having a small kinetic momentum $p$, such waves can reach the grid boundary, and the corresponding functions derivatives are small, so the Neumann boundary conditions appear to be more exact. 

\begin{figure}
\begin{center}
\includegraphics[width=0.5\textwidth]{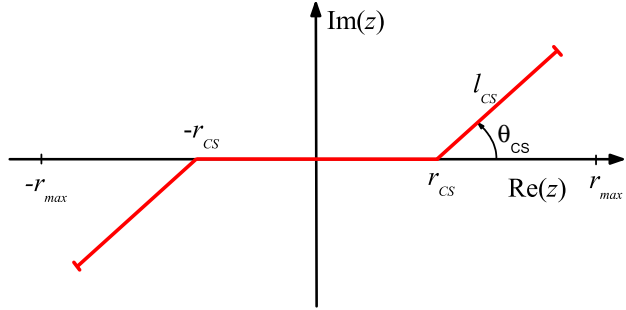}
\end{center}
\caption{(Color online) The complex contour $z$ introduced in the context of the ECS in the one-dimensional case.\label{FIG:ECS}}
\end{figure}

In the one-dimensional case one should use the contour with two scaling regions characterized by the two opposite values of the angles of the rotation to the complex plane, as represented in the Fig. \ref{FIG:ECS}.

The ECS technique is quite efficiently applied for the enforcing of the correct outgoing wave boundary conditions for the stationary scattering problems \cite{McCurdyRev2004}. But as for the time-dependent problems, the ECS runs into difficulties under the external field representation in the acceleration gauge. That is, in the variable external electric field the slow free electrons oscillate, thus periodically changing their kinetic momentum $\mathbf{p}(t)$ direction. Then under the electron ``backward motion'' the exponential function in the scaling regions becomes converted from decaying into the infinitely growing form, resulting in the appearance at the boundary of the huge value instead of the zero, therefore the complex scaling enhances the reflection instead of its suppression. This effect may be called the stimulated unphysical reflection. However, this challenge can be easily overcome by means of the choosing the scaling region size $l_{CS}$ being larger than the twice maximal amplitude of the electron oscillations driven by the external variable field. Then the stimulated unphysical reflected wave has no time enough to reach the real coordinate region $r<r_{CS}$ before the beginning of the next phase of the external field cycle and consequently the change of the $\mathbf{p}(t)$ direction to the original one.

\section{Suggested approach for amplitudes exctraction}\label{Sec:AmplExtr}

Now we turn to the exposition of the essence of the proposed approach for ionization amplitudes exctraction.

If the external field $\mathbf{E}(t)\neq 0$ only for a time period $0<t<T_{\text{pulse}}$, then the probabilty amplitude for the ejection of an electron having the momentum $\mathbf{p}$ may be expressed as
\begin{eqnarray}
A(\mathbf{k})=\left.\int\Phi_{\mathbf{k}}^*(\mathbf{r},t)\psi(\mathbf{r},t)d\mathbf{r}\right|_{t>T_{\text{pulse}}}.
\end{eqnarray}
Here $\Phi_{\mathbf{k}}(\mathbf{r},t)=\varphi_{\mathbf{k}}^{(-)}(\mathbf{r})e^{-\imath Et}$, $E=p^2/2$, $\varphi_{\mathbf{k}}^{(-)}$ is the eigenfunction of the unperturbed Hamiltonian
\[\hat{H}_0\varphi_{\mathbf{k}}^{(-)}(\mathbf{r})=E\varphi_{\mathbf{k}}^{(-)}(\mathbf{r}).\] From the amplitude one can obtain the probability density of the ionization implying the ejection of an electron having the momentum $\mathbf{k}$ as follows:
\begin{eqnarray}
\rho(\mathbf{k})=\frac{dP}{d\mathbf{k}}(\mathbf{k})=|A(\mathbf{k})|^2.\label{espectrum}
\end{eqnarray}
These expressions are applicable so long as the stationary wave functions $\varphi_{\mathbf{k}}(\mathbf{r})$ are known. Moreover, when any methods of the unphysical boundary reflections suppression are employed, the formula for the ejection amplitude becomes invalid because of the wavefunction departure from the TDSE solving region.

The authors of the work \cite{Selin1999} suggested an approach for the ionization probability amplitudes calculation not requiring for the knowing of the exact continuum function. It is based upon the Fourier transform of the time-dependent probability amplitude flux through a certain closed surface
\begin{eqnarray}
A(\mathbf{k})=\int_{0}^{\infty}\oint_{S} \mathbf{j}\left[\psi(\mathbf{r},t),\chi_{\mathbf{k}}^{(-)}(\mathbf{r},t)\right]\cdot\mathbf{n}_SdSdt. \label{IonizationAmplitude}
\end{eqnarray}
Here $S$ is the certain closed hypersurface surrounding the system and removed from the center far enough to neclect the system potential on it (usually a sphere with the radius $r_S$), $\mathbf{n}_S$ denotes its normal vector, and the testing function $\chi_{\mathbf{k}}^{(-)}(\mathbf{r},t)$ is an approximate TDSE solution for the space region outside $S$, describing an electron having the canonic momentum $\mathbf{k}$. Next, the probability amplitude flux vector is defined as
\begin{eqnarray}
\mathbf{j}[\psi,\varphi]=\frac{i}{2}[\psi\hat{\mathbf{D}}^*\varphi^*-\varphi^*\hat{\mathbf{D}}\psi],\label{flux}
\end{eqnarray}
where the operator $\hat{\mathbf{D}}=\nabla$ in the length gauge.
The testing function $\chi_{\mathbf{k}}^{(-)}(\mathbf{r},t)$ should satisfy the TDSE for $r>r_S$. When the system potential is short-range and negligible at $r>r_S$, the role of this function can be played by the TDSE solution for a free particle in a variable external electric field, i.e. the Volkov function \eqref{VolkovCoord}, $\chi_{\mathbf{k}}^{(-)}(\mathbf{r},t)=\Theta_{\mathbf{k}\,}(\mathbf{r},t)$. 

The solution of \eqref{TDSE3D} may be obtained through the numerical calculations only for a finite time period $t\in[0,T]$. For the aim of the ionization amplitude extraction one can use a formula equivalent to \eqref{IonizationAmplitude}, but implying the integration, as well, over a finite time period: 
\begin{eqnarray}
A_T(\mathbf{k})=\int_{0}^{T}\oint_{S} \mathbf{j}\left[\psi(\mathbf{r},t),\chi_{\mathbf{k}}^{(-)}(\mathbf{r},t)\right]\cdot\mathbf{n}_SdSdt.
 \label{AmplitudeSelin}
\end{eqnarray}
According to \cite{Tao2012}, we will refer to this approach as the time-dependent surface flux (t-SURFF) method. With the object of the expression \eqref{AmplitudeSelin} applicability, the electron after the end of the field action should have time enough to leave the sphere $S$, i.e. the condition $T\gg T_{\text{pulse}}+r_{S}/v$ ($v=p$ being the velocity of the electron which ejection amplitude is to be obtained) should be provided. If at rather large time value the wavefunction takes on the surface $S$ the value $\psi(\mathbf{r}_S,t>T)=0$, then $A(\mathbf{k})=A_T(\mathbf{k})$. But if the very-low-energy electrons have a non-negligible probability to be ejected, then $\psi(\mathbf{r}_S,t)$ converges to zero very slowly. In turn, the non-zero $\psi(\mathbf{r}_S,T)$ implies the results yielded by the formula \eqref{AmplitudeSelin} to oscillate with $T$ increasing (as is seen from the t-SURFF curves in the Fig. \ref{FIG:PeET}). In the work \cite{Serov2001} these oscillations were eliminated by means of the $A_T(\mathbf{k})$ averaging over $T$.

However, use can be made from the features both of the wavefunction possibility to be expressed after the end of the laser pulse through the unperturbed system propagator as
\begin{eqnarray}
\psi(\mathbf{r},t>T)=\exp[-i(t-T)\hat{H}_0]\psi(\mathbf{r},T), \label{psi_after_pulse}
\end{eqnarray}
as well as of the testing function to contain only the phase factor evolving after the external field switching off, namely
\begin{eqnarray}
\chi_{\mathbf{k}}^{(-)}(\mathbf{r},t>T)=\exp[-iE(t-T)]\chi_{\mathbf{k}}^{(-)}(\mathbf{r},T), \label{Volkov_after_pulse}
\end{eqnarray}
where $E=k^2/2$. The expression \eqref{IonizationAmplitude} might be transformed by the dividing of the time integral into the two constituent integrals, namely the one from 0 to $T$, and the other from $T$ to $\infty$, in the following way:
\begin{eqnarray}
A(\mathbf{k})=A_T(\mathbf{k})+A_B(\mathbf{k}). \label{AmplitudeSelinMcCurdy}
\end{eqnarray}
Here
\begin{eqnarray}
A_B(\mathbf{k})=\int_{T}^{\infty}\oint_{S} \mathbf{j}\left[\psi(\mathbf{r},t),\chi_{\mathbf{k}}^{(-)}(\mathbf{r},t)\right]\cdot\mathbf{n}_SdSdt
 \label{AmplitudeBeyond}
\end{eqnarray}
is the contribution to \eqref{IonizationAmplitude} from the amplitude probability flux which is to pass through the surface $S$ beyond the moment $T$. Using \eqref{psi_after_pulse} as well as \eqref{Volkov_after_pulse}, one can according to \cite{McCurdy2007}, formally take the integral over time in order to arrive to
\begin{eqnarray}
A_B(\mathbf{k})=\oint_{S}
\mathbf{j} \left[\psi_{sc}(\mathbf{r};T),\chi_{\mathbf{k}}^{(-)}(\mathbf{r},T)\right]\cdot\mathbf{n}_SdS,
 \label{AmplitudeMcCurdy}
\end{eqnarray}
where the function $\psi_{sc}(\mathbf{r};T)$ is determined from the relation
\begin{eqnarray}
\psi_{sc}(\mathbf{r};T)\equiv-\frac{1}{i\left(E-\hat{H}_0+i0\right)}\psi(\mathbf{r},T).
\end{eqnarray}
For the aim of $\psi_{sc}(\mathbf{r};T)$ obtaining one needs to solve the driven stationary Schrodinger equation
\begin{eqnarray}
\left(E-\hat{H}_0\right)\psi_{sc}(\mathbf{r};T)=i\psi(\mathbf{r},T)
 \label{psi_sc}
\end{eqnarray}
with the outgoing wave boundary condition. The equations (\ref{AmplitudeMcCurdy},\ref{psi_sc}) were derived in \cite{McCurdy2007}. The corresponding approach \cite{McCurdy2007,McCurdyRev2013} is based upon the supposing of the probability flux neglection at $t<T=T_{\text{pulse}}$ for the rather large $r_S$, that means therefore $A=A_B$. In what follows we will refer to the approach \cite{McCurdy2007} as the energy-dependent surface flux (E-SURFF) method. The approach proposed here and built upon the expression \eqref{AmplitudeSelinMcCurdy} (i.e. upon the combining of the results of the  t-SURFF and the E-SURFF methods), will be hereinafter referred to as the t\&E-SURFF. 

In the work \cite{Serov2011} we have suggested the approximate approach for the evaluation of the correction to \eqref{AmplitudeSelin}, based upon the following assumption: 
\begin{eqnarray}
\psi(\mathbf{r}_S,t>T)\simeq
\psi(\mathbf{r}_S,T)\exp[-i E_{eff}(t-T)].\label{psi_exp_t_asympt}
\end{eqnarray}
Here $E_{eff}$ is the complex ``effective energy'' computed by using of the formula
\begin{eqnarray} E_{eff}(\mathbf{r}_S,T)=\frac{i}{\psi(\mathbf{r}_S,T)}\left.\frac{\partial \psi(\mathbf{r}_S,t)}{\partial t}\right|_{t=T}. \label{E_eff}
\end{eqnarray}
This assumption is rest upon the wavefunction feature to approach at the boundary $\psi(\mathbf{r}_S,t)$ the superposition of the states with closely adjacent energies (approximately tending to the $E_{eff}$ real part) at large $t$, as well as its gradual decreasing caused by the persistent particles leaving. The $\psi(\mathbf{r}_S,t)$ descreasing is determined by the $E_{eff}$ image part, that results is the inequality $\Im E_{eff}<0$. Then \eqref{AmplitudeBeyond} evolves into 
\begin{eqnarray}
\tilde{A}_B(\mathbf{k})=-\oint_{S}\frac{1}{\imath\left(E-E_{eff}\right)}
\mathbf{j}\left[\psi(\mathbf{r},T),\chi_{\mathbf{k}}^{(-)}(\mathbf{r},T)\right]\cdot\mathbf{n}_S
dS. \label{IonizationAmplitudeCorr}
\end{eqnarray}
Finally one arrives to the formula 
\begin{eqnarray}
A(\mathbf{k})=A_T(\mathbf{k})+\tilde{A}_B(\mathbf{k}). \label{AmplitudeSerov}
\end{eqnarray}
This approach, despite demanding the solution for a time $T> T_{\text{pulse}}+r_{S}/v$ enough for the considered electron having the velocity $v$ to leave outside the $S$, has the advantages of both the relative ease of the expressions (\ref{E_eff},\ref{IonizationAmplitudeCorr}) using, as well as the eventual reduction of the required $T$ in comparison with the original t-SURFF utilizing (\ref{AmplitudeSelin}). In what follows we will refer to the approach based upon the expression \eqref{AmplitudeSerov}, as the corrected t-SURFF (t-SURFFc). 

In the presence of the Coulomb (i.e. non-short-range) system potential the Volkov functions become inapproproate for the testing functions $\chi_{\mathbf{k}}^{(-)}$ even for a large distance $r_S$ from the system center. For this case the authors of \cite{Selin2000} successfully used the WKB continuum wave-functions as the testing functions.

\section{Results of benchmark calculations}\label{Sec:NumericalModel}

\begin{figure}
\begin{center}
(a)\includegraphics[angle=270,width=0.5\textwidth]{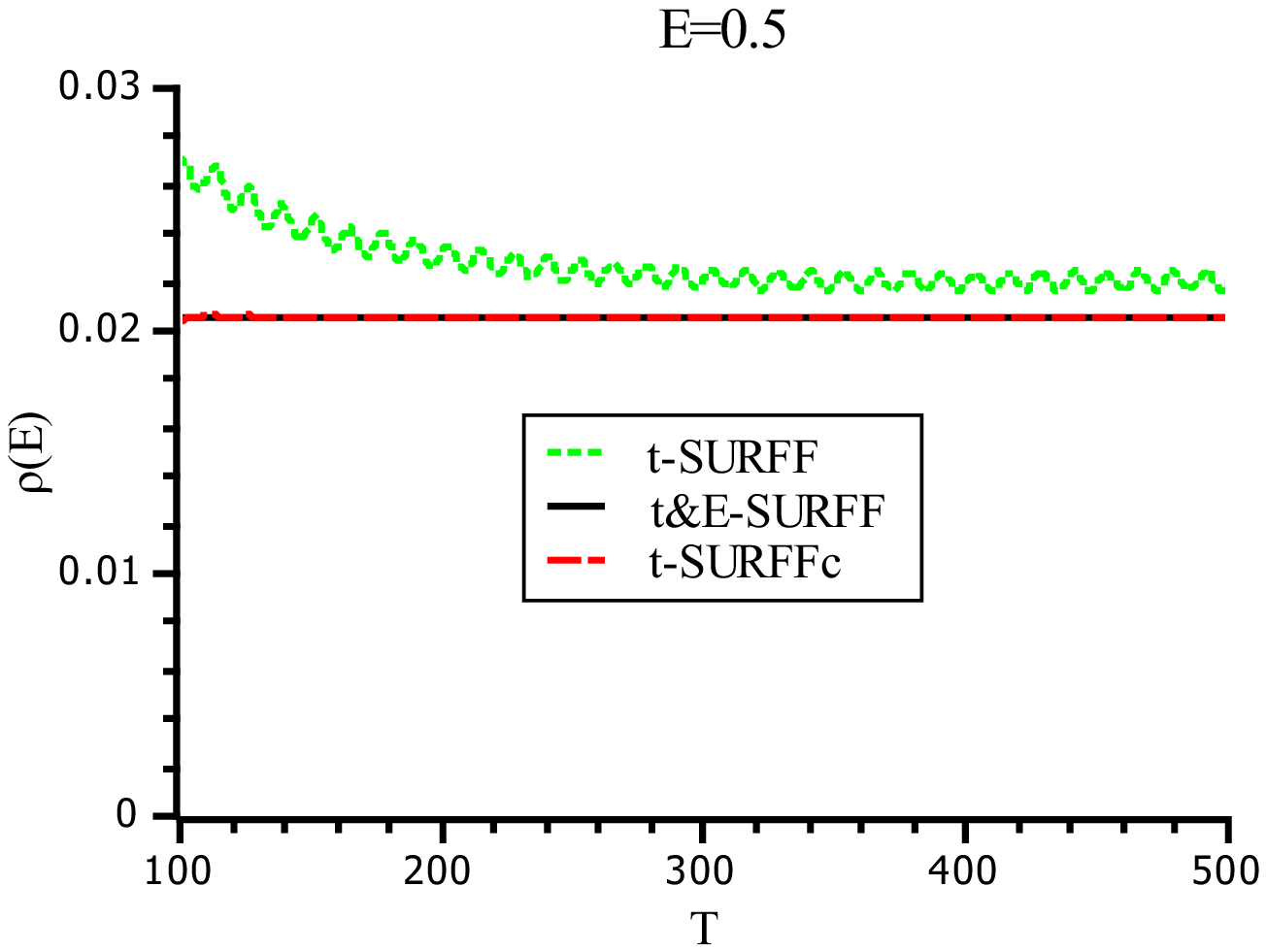}
(b)\includegraphics[angle=270,width=0.5\textwidth]{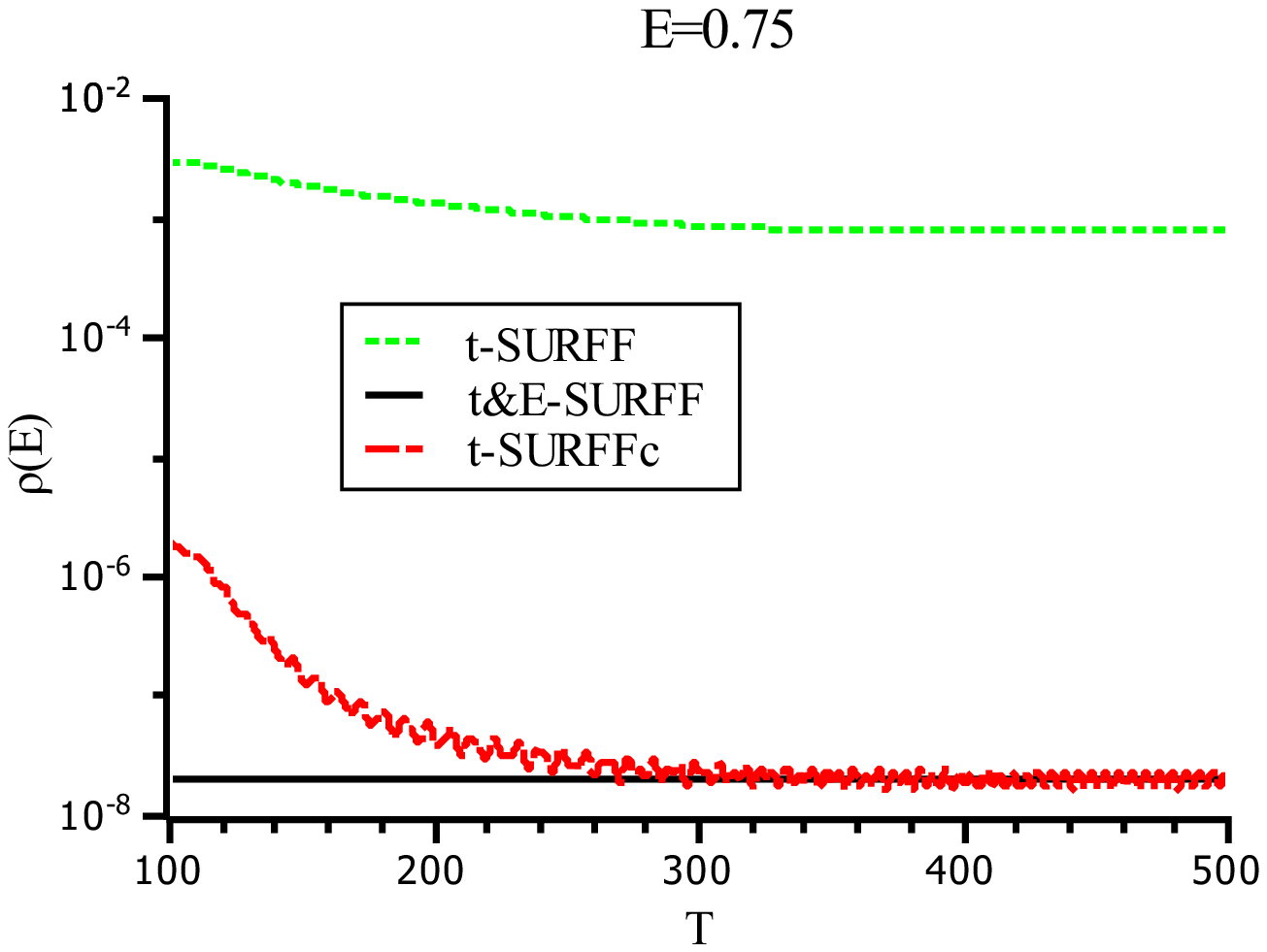}
(c)\includegraphics[angle=270,width=0.5\textwidth]{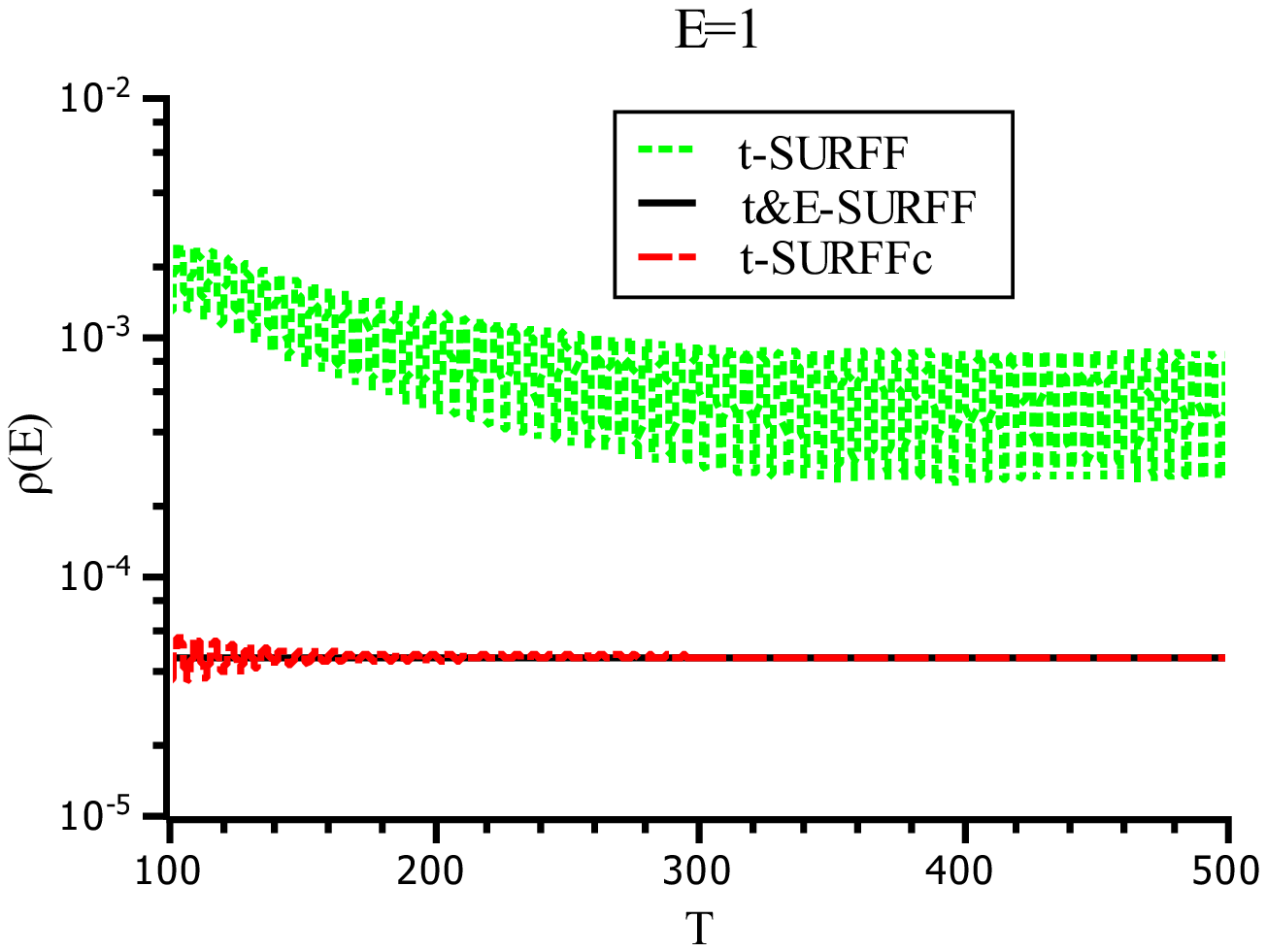}
\end{center}
\caption{(Color online) Ejected electron energy probability density $\rho(E)$ for fixed ejection energy $E$ as a function of integration time $T$, obtained by the different methods: t\&E-SURFF (solid line), t-SURFFc (dashed line), t-SURFF (dotted line). \label{FIG:PeET}}
\end{figure}

In order to demonstrate our method efficiency, we have performed the benchmark calculations. For this purpose in the capacity of a model system we used the one-dimensional system \cite{BouckeKull1997,Selin1999}
\begin{eqnarray}
i\frac{\partial\psi(x,t)}{\partial
t}=\left[-\frac{1}{2}\frac{\partial^2}{\partial x^2}+U_0(x)-\mathcal{E}(t)x\right]\psi(x,t)
\label{Shr1D}
\end{eqnarray}
with the P\"oschl--Teller potential
\[
U_0(x)=-\frac{1}{\cosh^2 x},
\]
in the variable external field
\[
\mathcal{E}(t)=\mathcal{E}_0\sin^2(\pi t/T_{\text{pulse}})\sin\omega t
\]
The system with such potential has the only one bound state 
\[
\psi(x,0)=\psi_0(x)=\frac{1}{\sqrt{2}\cosh x},
\]
with the corresponding energy $E_0=-0.5$, and it was just taken as an initial state. In all the cases considered the frequency was set to $\omega=0.5$ (which value coincides with the system ionization potential $I=|E_0|=0.5$), the field strength amplitude to $\mathcal{E}_0=0.05$, and the laser pulse duration to $T_{\text{pulse}}=8T_{osc}$, $T_\text{osc}=2\pi/\omega$. The frequency adjustment to the transition from the ground state to the zero energy state results in the significant number of the very-low-energy electrons leaving the calculation region very slowly, therefore providing a stringent test of the methods under comparison. The one-dimensional Volkov function was used as the testing function
\begin{eqnarray}
\chi_{k}(x,t)=\frac{1}{\sqrt{2\pi}}
\exp\left\{i\left[
p(t)x-\frac{1}{2}\int_0^t p^2(\tau)d\tau
\right]\right\}. \label{VolkovCoord1D}
\end{eqnarray} 

For the aim of the Eq. \eqref{Shr1D} numerical solution we used the simplest numerical scheme resting upon the 2nd order finite difference formula for the second derivative over $x$, and the Cranck-Nickolson scheme for the time evolution as well. The Eq.\eqref{psi_sc}, whose numerical solution is required for both the E-SURFF and t\&E-SURFF implementation, was solved with the help of the LU-decomposition method for the tridiagonal matrix. The grid steps $h=0.01$ and $dt=0.01$ were selected small enough to both the space and time approximation errors to be insignificant in comparison with other sources of errors. 

For the one-dimensional problem a closed surface $S$ having the hyperradius $r_S$ reduces to the two points $x_S=\pm r_S$ on the $x$ axis. However, easy to show that the contribution of the probability amplitude flux through the point opposite to the electron ejection direction to the result of Eq.\eqref{IonizationAmplitude} equil zero. But for the finite time integrals entering the \eqref{AmplitudeSelin} and \eqref{AmplitudeMcCurdy} relations this term leads to the strong extra oscillations over energy in the corresponding methods results. Their summation that is implied in t\&E-SURFF makes the oscillating term disappear, so that the one arrives to the same result as obtained by means of the flux through the only one point. In other words, the using of the flux through the both two points yields only slowing of the t-SURFF and E-SURFF convergence. From this reason, here for the evaluation of the probability amplitude of the electron ejection in the left direction (i.e. for $k<0$) we take into account only the flux through the left point $x_S=-r_S$ only, as well as the flux through the right point $x_S=r_S$ alone for the corresponding case of ejection to the right direction ($k>0$).

Next, the complex contour has the shape as presented in the Fig.\ref{FIG:ECS}. The scaling points of rotation to the complex plane $x_{CS}=\pm r_{CS}$ for all the calculations were selected in the vicinity of the amplitudes extraction point $r_{CS}=r_S+h$. The length of the complex scaled contour part was selected to be $l_{CS}=10$, and the scaling rotation angle to $\theta_{CS}=45^\circ$, thus providing $\max{\Im x}=10/\sqrt{2}$.

\begin{figure}
\begin{center}
\includegraphics[angle=270,width=0.5\textwidth]{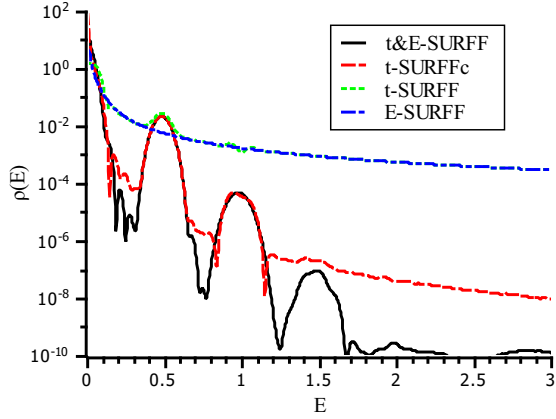}
\caption{(Color online) The $\rho(E)$ as a function of the electron energy $E$, extracted from the single $\psi(x,t)$ by various methods. $x_S=10$, $T=8T_\text{osc}$. Results of t\&E-SURFF (solid line), t-SURFFc (dashed line), t-SURFF (dotted line), and E-SURFF (dash-dot line).\label{FIG:PeT8}}
\end{center}
\end{figure}

The Fig. \ref{FIG:PeT8} presents the probability density $\rho(E)$ of the ejection of the electron having the energy $E$
\[
\rho(E)=\frac{dP}{dE}=\frac{1}{k}\left[|A(-k)|^2+|A(k)|^2\right],
\] 
evaluated by the methods t\&E-SURFF, t-SURFFc, t-SURFF, and E-SURFF at one and the same exctraction point $x_S=10$ (and therefore the same space grid) and the same time period $T=T_{\text{pulse}}$ as well. 

The Fig. \ref{FIG:PeET} demonstrates the  $\rho(E)$ convergence with the increasing of $T$ ($T>T_{\text{pulse}}$). The energy values, which the $\rho(E)$ displayed in the Figs. \ref{FIG:PeET}a and \ref{FIG:PeET}c were calculated for, correspond to the maxima of the t\&E-SURFF curve, and that for the Fig. \ref{FIG:PeET}b --- to the minimum. It is clearly seen that the t\&E-SURFF results depend by no means on $T$, besides the t-SURFFc results converge quite rapidly to the ones yielded by the t\&E-SURFF, unlike the original t-SURFF that apparently converges very slowly and suffers strong oscillations. Its most likely reason is the appearance of a significant number of the low-energy electrons at the chosen external field parameters and consequently the extremely slow wavefunction decreasing in the amplitudes extraction points. Oscillations appear in the t-SURFF results because of the addition to the true ionization amplitudes of the extra parasite term emerging due to a non-zero $\psi(x_S,t)$ meaning at $t\geq T$. One should also mention no oscillations in the t-SURFF curve in the Fig. \ref{FIG:PeET}b; this can be explained by the fact that since the $\rho(E)$ has a deep minimum at the energy $E=0.75$, the true amplitude becomes negligible relatively to the parasite term, so that the dependence obtained by the t-SURFF contains only the parasite term eventually decreasing with time because of the $\psi(x_S,t)$ decay.

\begin{figure}
\begin{center}
\includegraphics[angle=270,width=0.5\textwidth]{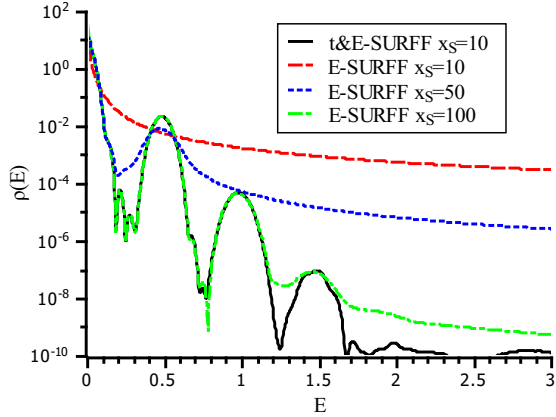}
\caption{(Color online) 
The $\rho(E)$ for $T=T_{\text{pulse}}$ extracted by the t\&E-SURFF for $r_S=10$ (solid line), and by the E-SURFF for $r_S=10$ (dashed line), $r_S=50$ (dotted line), and $r_S=100$ (dash-dot line). \label{FIG:ESURFFconv}}
\end{center}
\end{figure}

Next, the Fig. \ref{FIG:ESURFFconv} illustrates the convergence of the E-SURFF results with the increasing of the distance $r_S$ to the extraction points. One can observe the spectrum obtained by means of the E-SURFF convergence with the $r_S$ growth to the one yielded by the t\&E-SURFF for $r_S=10$, at that achieved at the small energies. The reason is obviously the fact that the E-SURFF implies the amplitudes extracting from the wavepacket in the region bounded by $r_S$, and the main contribution to the wave packet near the center comes from the low-energy electrons.

In sum, the t\&E-SURFF method appears to be the most efficient approach from the ones considered here, whereas it allows to obtain the probability amplitudes by means of the TDSE solution for rather moderate both space region as well as integration time. The t-SURFFc method, though not providing such a rapid time convergence, appears to be much more efficient as compared to the original t-SURFF, besides it does not require the solving of Eq.\eqref{psi_sc} unlike the t\&E-SURFF.

\section{Conclusion}\label{Sec:Concl}

In sum, in the present work we propose the new approach for the ionization amplitudes exctraction from the evaluated wavefunction without necessity of knowing of the exact continuum function. The method, designated the t\&E-SURFF, consistes in the composition of the results obtained with using of the t-SURFF \cite{Selin1999,Tao2012} and E-SURFF \cite{McCurdy2007} techniques. 

The t-SURFF method actually yields the probability amplitude for a particle having the momentum $\mathbf{k}$ to leave a volume region bounded by a surface $S$ \textit{before} $t=T$. This technique allows the amplitudes obtaining via the TDSE solving that should be performed though in rather small space grid, but for a time period much exceeding the external filed action duration in order for the particles to reach the boundary $S$ as well as for the wavefunction to turn into zero on this surface. At the same time, the E-SURFF approach yields the probability amplitude for the particle having the momentum $\mathbf{k}$ to leave the volume region bounded by the surface $S$ \textit{after} the moment $t=T$. This method implementation needs the TDSE solving though only for a time period corresponding to non-zero ionizing field value, but for the enormous space region $S$ that should enclose the whole wave packet after the end of the external action. The t\&E-SURFF approach allows to combine the advantages of the both methods since it implies the TDSE solving for both small space region as well as rather short time period $T=T_{\text{pulse}}$ of the external field action to the system.

The approach we previously proposed \cite{Serov2011}, namely the t-SURFFc, appears to be easier to use by contrast to the t\&E-SURFF and generally demands the less $T$ value as compared to the original t-SURFF, though it still should meet the condition $T>T_{\text{pulse}}$. However, the t-SURFFc in its simplest form can be applied only for the one-body problems. On the contrary, in the two-body problems such as the helium double photoionization \cite{Serov2007} the wavefunction on the boundary is revealed as the superposition of the states describing either the double photoionization as well as the single ionization with one electron remained bound in a number of the residual ion excited states. The latter have substantionally different energies, as a result $|\psi(\mathbf{r}_S,t)|$ even at very large $t$ values does not tend monotonically to zero, but instead oscillates in a complicated fashion, so that the approximation \eqref{psi_exp_t_asympt} becomes invalid, and  under the effective energy evaluation via the relation \eqref{E_eff} the $\Im E_{eff}$ periodically takes the positive values. Among other things, this particularity complelled us after the attempts of the t-SURFFc utilizing to employ \cite{Serov2007} the  approach for the double ionization amplitude exctraction. 

Since both the t-SURFF and E-SURFF methods utilize the same expression for the probability amplitude flux through the boundary, the logarithmic phase addition emerging from the interelectron correlation presence \cite{McCurdy2003} appears to be the same for these two methods therefore allowing to combine easily their results thus realizing the t\&E-SURFF approach for the numerical description of the double ionization. For the aim of the two-electron system single ionization evaluation by means of the t-SURFF technique the author of \cite{Scrinzi2012} has proposed the effective testing fuctions obtained via the backward time propagation of the ion bound states. The combining of the approach suggested in that work with the E-SURFF is also trivial. Nevertheless, the examination of the effectivity of t\&E-SURFF for the description of the one- and manyfold ionization of a many-particle system seems to be a matter to a separate dedicated study.

\acknowledgments The authors acknowledge support of the work from the Russian Foundation for Basic Research (Grant No. 11-01-00523-a).

\appendix
\section{The surface flux for the different external field gauges}
In the above consideration the external field was treated in a length gauge. The latter yield a benefit of the ability to describe the variable external field through a scalar potential. Consequently, the external field operator in the position representation becomes diagonal that appears to be very useful when employing the discrete variable representation in combination with the split-operator method (like as used in \cite{Serov2011}). However, another authors (e.g., \cite{Selin1999} and \cite{Tao2012}) utilize the velocity gauge where the Hamiltonian of a system under the action of the variable external field has the form 
\begin{eqnarray}
\hat{H}_v(\mathbf{r},t)=\frac{1}{2}\left[-i\nabla-\frac{e}{c}\mathbf{A}(t)\right]^2 + U_0(\mathbf{r}).
\end{eqnarray}
Also, in some cases it appears to be more convenient to use the acceleration gauge (see e.g. \cite{BouckeKull1997}), transform to which might be performed by means of the Kramers-Henneberger transformation with the subtitution $\mathbf{r}=\mathbf{r}'+\mathbf{r}_e(t)$. In this frame the Hamiltonian takes the form
\begin{eqnarray}
\hat{H}_a(\mathbf{r}',t)=-\frac{1}{2}\nabla_{\mathbf{r}'}^2 + U_0[\mathbf{r}'+\mathbf{r}_e(t)].
\end{eqnarray}
where $\mathbf{r}_e(t)=-\frac{e}{c}\int_0^t\mathbf{A}(t')dt'$ is the law of motion for a classical particle in the external field, $\mathbf{r}'=\mathbf{r}-\mathbf{r}_e(t)$ is the transformed coordinate vector. 

The using of the different gauges implies the certain distinction in the details of the probability amplitude flux evaluation. That is, first, the Volkov function takes distinct forms in different gauges. Its length form is given in the Eq. \eqref{VolkovCoord}. Next, in the velocity gauge it is presented as
\begin{eqnarray}
\Theta_{\mathbf{k}\,}(\mathbf{r},t)=
(2\pi)^{-3/2} \exp\left\{i\left[\mathbf{k}\cdot\mathbf{r}-\frac{k^2}{2}t\right]\right\},
 \label{VolkovVeloc}
\end{eqnarray}
Finally, in the acceleration gauge the Volkov function is represented simply as a plane wave
\begin{eqnarray}
\Theta_{\mathbf{k}\,}(\mathbf{r}',t)=(2\pi)^{-3/2} \exp\left[i\left(\mathbf{k}\cdot\mathbf{r}'-\frac{k^2}{2}t\right)\right].  \label{VolkovAccel}
\end{eqnarray}

Secondly, for the different gauges one has the distinct forms of the operator $\hat{\mathbf{D}}$ entering the flux definition \eqref{flux}. Namely, the operator $\hat{\mathbf{D}}$ is represented correspondingly in the length, velocity, and acceleration gauge as follows:
\begin{eqnarray}
 \hat{\mathbf{D}}_{\text{c}}&=&\nabla;\\
 \hat{\mathbf{D}}_{\text{v}}&=&\nabla-i\frac{e}{c}\mathbf{A}(t);\\
 \hat{\mathbf{D}}_{\text{a}}&=&\nabla_{\mathbf{r}\,'}.
\end{eqnarray}

\end{document}